\documentclass{aastex}
\usepackage{emulateapj5}
\usepackage{lscape}
\usepackage{onecolfloat}   %%% If you omit this, then remove the two lines
\newcounter{affil}
\newcommand{\hoggaffil}[2]{%
	\addtocounter{affil}{1}%
	\altaffiltext{\theaffil}{{#2}\label{#1}}}

\newcommand{\eg}{{\it e.g.}}

\newcommand{\hmpc}{\ensuremath{h^{-1}{\rm\,Mpc}}}
\newcommand{\ihmpc}{\ensuremath{h{\rm\,Mpc}^{-1}}}

\newcommand{\hgpcC}{\ensuremath{h^{-3}{\rm\,Gpc^3}}}
\newcommand{\ihmpcC}{\ensuremath{h^3 {\rm\,Mpc}^{-3}}}
\newcommand{\kmsmpc}{\ensuremath{{\rm\ km\ s^{-1}\ Mpc^{-1}}}}

\newcommand{\wrp}{{w_p(r_p)}}

\newcommand{\kmsMpc}{\kmsmpc}

\newcommand{\tableskip}{\\[-6pt]}
\newcommand{\singleline}{\tableskip\hline\tableskip}
\newcommand{\doubleline}{\tableskip\hline\hline\tableskip}

\newcommand{\outline}[1]{\relax}

\begin{document}
\twocolumn[%%% Begin one-column header material
\submitted{}
\lefthead{Clustering of Luminous Red Galaxies}

\title{The Intermediate-scale Clustering of Luminous Red Galaxies}
\author{
Idit Zehavi\altaffilmark{\ref{Arizona}}, 
Daniel J.\ Eisenstein\altaffilmark{\ref{Arizona},\ref{SF}},
Robert C.\ Nichol\altaffilmark{\ref{Portsmouth}}, 
Michael R.\ Blanton\altaffilmark{\ref{NYU}}, 
David W.\ Hogg\altaffilmark{\ref{NYU}}, 
Jon Brinkmann\altaffilmark{\ref{APO}},
Jon Loveday\altaffilmark{\ref{Sussex}},
Avery Meiksin\altaffilmark{\ref{Edinburgh}},
Donald P.\ Schneider\altaffilmark{\ref{PennState}}, \\
and Max Tegmark\altaffilmark{\ref{Penn}}
}

\begin{abstract}
We report the intermediate-scale ($0.3$ to $40 \hmpc$) clustering
of $35,000$ luminous early-type galaxies at redshifts $0.16$ to $0.44$ 
from the Sloan Digital Sky Survey.  We present the redshift-space 
two-point correlation function $\xi(s)$, the projected correlation function 
$\wrp$, and the deprojected real-space correlation function $\xi(r)$,
for approximately volume-limited samples.
As expected, the galaxies are highly clustered, with 
the correlation length varying from $9.8 \pm 0.2 \hmpc$ to 
$11.2 \pm 0.2 \hmpc$, dependent on the specific luminosity range.
For the $-23.2<M_g<-21.2$ sample, 
the inferred bias relative to that of $L_*$ galaxies is  
$1.84 \pm 0.11$ for $1\hmpc < r_p \la 10\hmpc$,
with yet stronger clustering on smaller scales.
We detect luminosity-dependent bias within the sample but see 
no evidence for redshift evolution between $z=0.2$ and $z=0.4$.
We find a clear indication for deviations from a power-law in the 
real-space correlation
function, with a dip at $\sim\!2 \hmpc$ scales and an upturn on smaller scales.
The precision measurements of these clustering trends offer new avenues
for the study of the formation and evolution of these massive galaxies.
\end{abstract}

]%%% end twocolumn header

\hoggaffil{Arizona}{Steward Observatory, University of Arizona,
		933 N. Cherry Ave., Tucson, AZ 85121}
\hoggaffil{Portsmouth}{Institute of Cosmology and Gravitation, 
    Mercantile House, Hampshire Terrace, University of Portsmouth,
    Portsmouth, P01 2EG, UK}               
\hoggaffil{NYU}{Center for Cosmology and Particle Physics, 
    Department of Physics, New York University,
    4 Washington Place, New York, NY 10003}
\hoggaffil{APO}{Apache Point Observatory,
                 P.O. Box 59, Sunspot, NM 88349}
\hoggaffil{Sussex}{Astronomy Centre, 
		University of Sussex, Falmer, Brighton BN1 9QJ, UK}
\hoggaffil{Edinburgh}{Institute of Astronomy, University of Edinburgh,
           Blackford Hill, Edinburgh, EH9 3HJ, UK}
\hoggaffil{PennState}{Department of Astronomy and Astrophysics,
                 Pennsylvania State University, University Park, PA 16802}
\hoggaffil{Penn}{Department of Physics, University of Pennsylvania, 
           Philadelphia, PA 19104}
\hoggaffil{SF}{Alfred P.~Sloan Fellow}

%\vspace{0.4cm}
\section{Introduction}
\label{sec:intro}

The clustering of galaxies provides a window not only to the formation
of inhomogeneities in the early universe but also onto the physics of
galaxy formation.  Galaxies with different properties cluster differently
\citep{hubble36,zwicky68,davis76,dressler80,postman84,hamilton88,white88,park94,loveday95,guzzo97,benoist96,Wil98,Bro00,Car01,Nor01,Zeh02,Nor02,Bud03,madgwick03,Hog03,Zeh04b},
and these trends can be connected to their
small-scale environments, notably the masses of their host dark matter
halos (\eg, \citealt{Kai84,BBKS,Mo96,Ben00,She01,Ber03}).  This path has 
been strengthened recently by the discovery of
deviations from the canonical power-law correlation function on small
scales (\eg, \citealt{Zeh04a,Zhe04}) and the ease of interpretation of these 
features by contemporary models of galaxy and halo clustering, in terms of 
the clustering of galaxies within single halos and the clustering between halos
\citep{Kau97,jing98,Kau99,Ben00,Ma00,Pea00,Sel00,Sco01,Ber02,Ber03,Mag03,Kra04,Zeh04a,Zhe04}.

The Sloan Digital Sky Survey (SDSS;\citealt{Yor00}) was designed in scope 
and systematic control to permit the study of galaxy clustering over a wide 
range of scales and galaxy properties 
(\eg, \citealt{Con02,Zeh02,Bud03,Hog03,Teg04,Zeh04b};
all using the SDSS main galaxy sample). 
To improve the precision of clustering measurements on the largest scales, 
the SDSS provides a spectroscopic sample of luminous red galaxies (LRG).
These galaxies reach a redshift of 0.5, thereby providing a sample of over 
$1\hgpcC$ (see \citealt{Eis01}). 
Thus far, over 50,000 spectra of LRGs have been acquired.

In this paper, we will investigate the clustering of these luminous early-type
galaxies on scales between $0.3$ and $40\hmpc$.  This stretches from
the quasi-linear to the deeply non-linear regime.  As massive early-type
galaxies are known to inhabit preferentially rich environments (\eg,
\citealt{San72,dressler80,Hoe80,Sch83,postman84,Pos95},M.\ Bernardi 2004, in
preparation), this
selection should permit one to study the clustering and internal
structure of massive halos.  Models that differ in their association of
LRGs to cluster-sized halos or to the fraction in smaller halos will
vary not only in their predicted correlation length, but also in the
fine structure of the correlation functions.
With the sample size available
within the SDSS LRG sample, we expect to reach the precision necessary to
perform such tests despite the rarity of massive galaxies.
We note that the rapid increase in the clustering
of early-type galaxies at the highest luminosities \citep{Hog03,Zeh04b}
implies that the connections
between the most massive galaxies and their environments is notably
different than even $L_*$ early-types.

The outline of the paper is as follows. 
In \S~\ref{sec:data} we present the LRG sample. 
In \S~\ref{sec:sclustering} we present the clustering measurements in 
redshift-space and in \S~\ref{sec:rclustering} we show the inferred 
real-space clustering results. We conclude in \S~\ref{sec:discussion}.
Details of our sample modeling are given in the Appendix.
 
Throughout the paper, all distances are comoving and quoted in $\hmpc$, 
where $h \equiv H_0/100 \kmsMpc$. For all distances and absolute magnitude 
we use a cosmology of $\Omega_m=0.3$ and $\Lambda=0.7$ and adopt $h=1$ to 
compute absolute magnitudes.

%\pagebreak

\section{Data}
\label{sec:data}

\subsection{The SDSS LRG Sample}
\label{subsec:data}

The SDSS \citep{Yor00} is imaging $10^4$ square degrees away from the Galactic Plane
in 5 passbands, $u$, $g$, $r$, $i$, and $z$ \citep{Fuk96,Gun98}.  
Image processing \citep{Lup01,Sto02,Pie03} and calibration
\citep{Hog01,Smi02} allow one to select galaxies, quasars, and stars
for follow-up spectroscopy with twin fiber-fed double-spectrographs.
The spectra cover 3800\AA\
to 9200\AA\ with a resolution of 1800.  Targets are assigned to plug
plates with a tiling algorithm that ensures nearly complete samples 
\citep{Bla03a}. An operational constraint of using fibers to obtain
spectra is that no two fibers can be closer than $55''$ on the same 
plate. This constraint is partly alleviated by having roughly a third
of the sky covered by overlapping plates. 

Galaxy spectroscopic target selection proceeds by two algorithms. 
The primary sample \citep{Str02}, referred to here as the MAIN sample,
targets galaxies brighter than $r<17.77$.   The surface density of such
galaxies is about 90 per square degree.  
The LRG algorithm \citep{Eis01} then selects $\sim\!12$ additional galaxies
per square degree, using color-magnitude cuts in $g$, $r$, and $i$
to select galaxies to $r<19.5$ that are likely to be luminous early-types
at redshifts up to $\sim\!0.5$.  The selection is extremely efficient,
and the redshift success rate is very high.  A few galaxies (3 per square degree
at $z>0.15$) matching the rest-frame color and luminosity properties of the LRGs 
are extracted from the MAIN sample; we refer to this combined set as the 
LRG sample.  In detail, there are two parts to the LRG algorithm, known
as Cut I and Cut II and described in \citet{Eis01}.

We begin from a spectroscopic sample covering 3,836 square degrees.  
The exact survey geometry is expressed in terms of spherical polygons 
and is known as {\tt lss\_sample14} (M. Blanton 2004, in preparation). 
This set contains 55,000 spectroscopic LRGs in the redshift range 
$0.15<z<0.55$.

\subsection{Redshift and Magnitude Cuts}
\label{subsec:samples}

\begin{figure}[tb]
\plotone{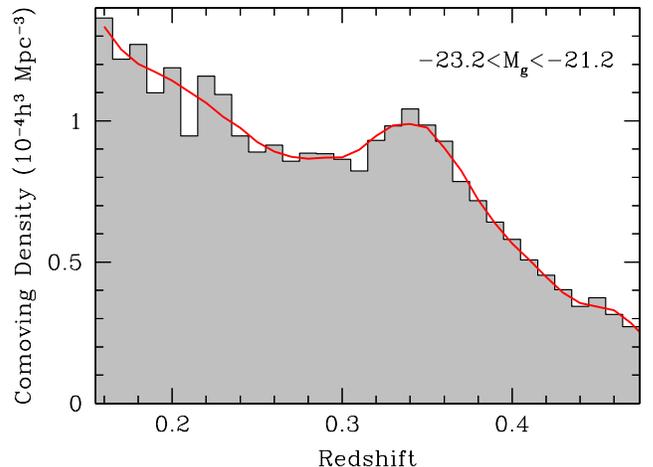}
\caption{\label{fig:nz210}
The comoving number density of the $-23.2<M_g<-21.2$ sample. The shaded
histogram is the distribution of the actual data, and the solid continuous
line is our model for the redshift distribution, described in Appendix A.
The sample is close to a constant comoving volume for $z<0.36$, although the
fluctuations are reaching about 30\% peak-to-peak (but one should note that the
lowest redshifts, where the excess is, contain less volume than the redshift 
range would suggest).
}
\end{figure}

\begin{figure}[tb]
\plotone{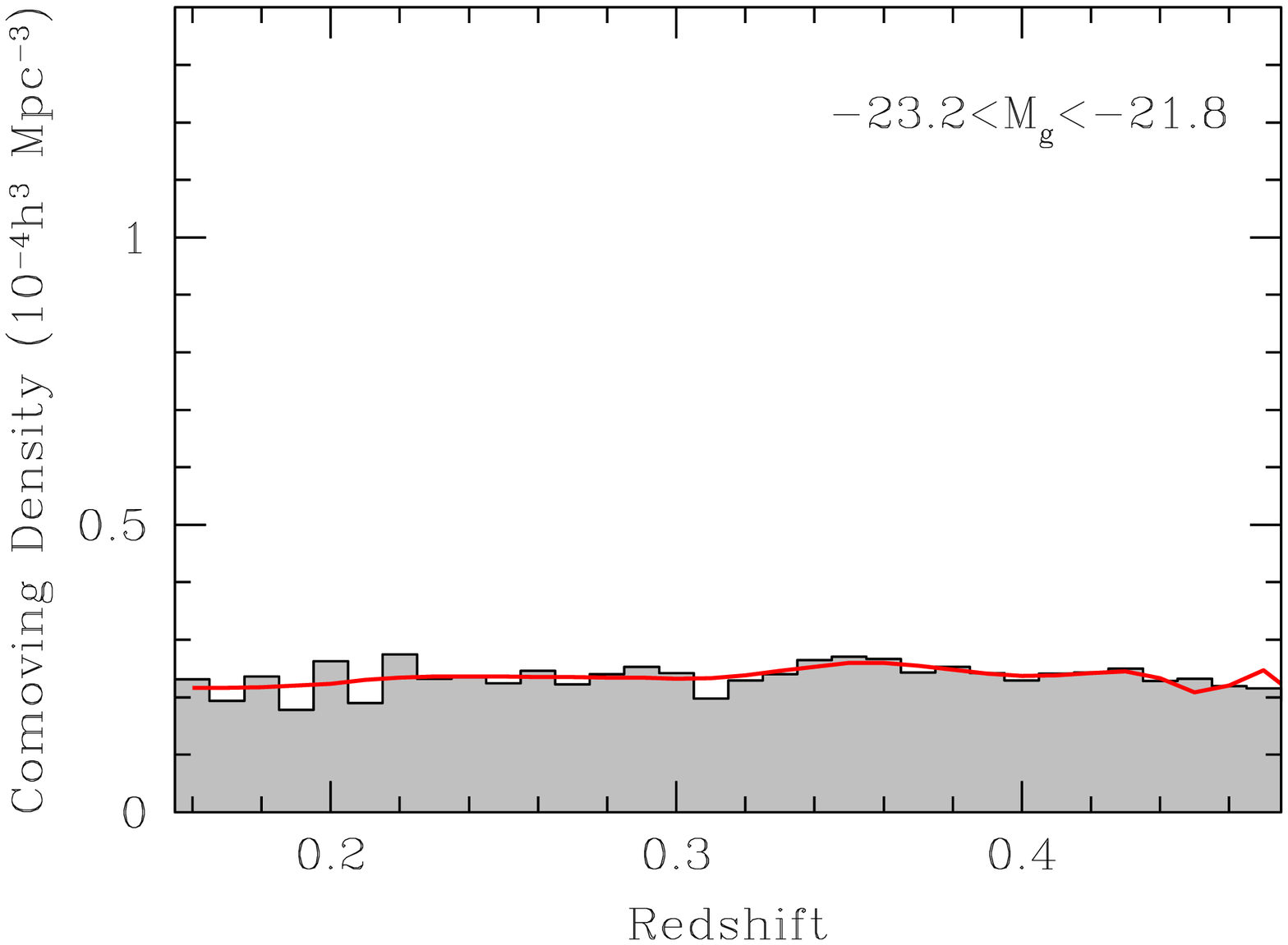}
\caption{\label{fig:nz216}
As Figure \protect\ref{fig:nz210}, but for the $-23.2<M_g<-21.8$ sample.
The sample is close to a constant comoving volume for $z<0.44$.
}
\end{figure}

\begin{figure}[tb]
\plotone{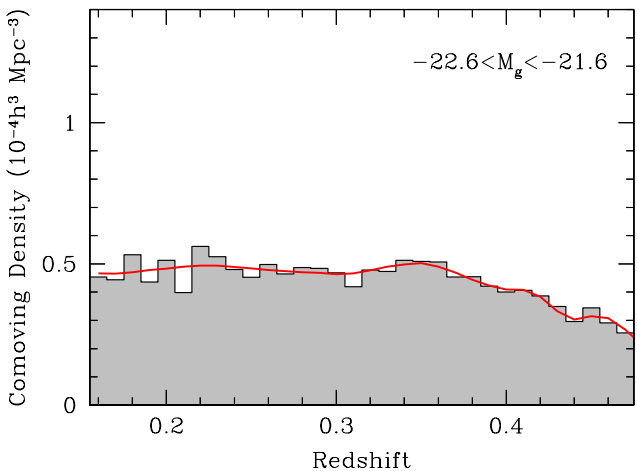}
\caption{\label{fig:nz214}
As Figure \protect\ref{fig:nz210}, but for the $-22.6<M_g<-21.6$ sample.
The sample is close to a constant comoving volume for $z<0.36$.
}
\end{figure}

The SDSS LRG sample is nearly volume-limited, but not precisely so.  
At $z>0.37$, the flux limits of $r<19.2$ (Cut I) and $r<19.5$ (Cut II)
begin to move into the passively-evolving luminosity threshold.  In this
paper, we wish to analyze volume-limited samples, 
so as to study the clustering properties of well defined populations of
galaxies. 
We therefore define three subsamples in passively-evolved luminosity and
restrict the redshift ranges to ensure complete coverage.  The subsamples
are $-23.2<M_g<-21.2$ with $0.16<z<0.36$, $-23.2<M_g<-21.8$ with $0.16<z<0.44$,
 and $-22.6<M_g<-21.6$ with $0.16<z<0.36$.
The first of these sets is picked to maximize our use of the LRG spectroscopy
for the innately volume-limited portion of the sample.  The second is
selected because the Cut II selection creates a knee in the number densities
as a function of redshift that we can exploit.  The third is chosen to match 
the luminosity range of the $-23<M_r<-22$ volume-limited MAIN galaxy sample 
described in \citet{Zeh04b}. We use here only the red subsample of the latter
(as defined in \citealt{Zeh04b}, with $0.10<z<0.23$) to compare to the LRG 
clustering. These LRG and MAIN samples overlap nearly completely in the 
redshift range in common, $0.16<z<0.23$.
The basic information regarding the three LRG samples is summarized in
Table~\ref{tab:samples}, and their comoving number density as a function
of redshift is shown in Figures~\ref{fig:nz210}-\ref{fig:nz214}. 

\begin{table*}[tb]\footnotesize
\caption{\label{tab:samples}}
\begin{center}
{\sc LRG Sample Statistics\\}
\begin{tabular}{cccccc}
\doubleline
$M_g$\tablenotemark{a} & Redshift & Number & Density\tablenotemark{b} & $\left<M_g\right>$\tablenotemark{c} & $\left<z\right>$\tablenotemark{d} \\
\singleline
$-23.2<M_g<-21.2$       & $0.16<z<0.36$ &  29298 & $9.7\times10^{-5}$ 
& -21.63  & 0.28 \\
$-23.2<M_g<-21.8$     & $0.16<z<0.44$ &  12992 & $2.4\times10^{-5}$ 
& -22.01 & 0.34 \\
$-22.6<M_g<-21.6$ & $0.16<z<0.36$ &  14500 & $4.8\times10^{-5}$ 
& -21.84 & 0.28 \\
\singleline
\end{tabular}
\end{center}
%NOTES.---%
\tablenotetext{a}{Rest-frame $g$-band absolute magnitudes, passively evolved 
to $z=0.3$.}
\tablenotetext{b}{Average comoving densities are in units of $\ihmpcC$.}
\tablenotetext{c}{Average rest-frame $g$-band absolute magnitude, $M_g$}
\tablenotetext{d}{Average redshift}
\end{table*}

The above luminosities have been $k$-corrected and passively evolved
to rest-frame magnitudes at $z=0.3$ (near the median redshift of the
LRG sample).  We use the observed $r$-band to estimate rest-frame $g$, 
as this requires minimal $k$-corrections at $z=0.3$.
We have used the ``non-star-forming'' model presented in Appendix B of \citet{Eis01}
but normalized to $M_g$ at $z=0.3$.
The model has relatively mild evolution, only about 1 magnitude per unit redshift,
compared to other measurements \citep{Bla03c}.  
Additional details of the samples' modeling are given in Appendix A. 
To the extent that our model is appropriate, our selections represent mass
thresholds throughout the sample volume.  However, as shown in Figure
\ref{fig:nz210}, the $-23.2<M_g<-21.2$ sample still
has some redshift evolution in the number density.
This is due to small fluctuations in the selection thresholds of the parent sample
in luminosity and rest-frame color as a function of redshift.
The other two samples, being safely more luminous than the LRG sample
selection limits, are much closer
to a constant comoving threshold (Figs.\ \ref{fig:nz216} and \ref{fig:nz214}).

\section{Redshift-Space Clustering}
\label{sec:sclustering}

We calculate the LRG correlation function in redshift space as a
function of the redshift-space separation $s$. To estimate the mean 
density and account for the complex survey geometry, we generate random 
catalogs, applying the 
radial and angular selection functions of the samples. The details of 
the radial and angular modeling are given in the Appendix.
We typically use in each random catalog 100-150 times the
number of galaxies in the real sample, and we have verified that changing
the random catalog makes negligible difference to the results. 
We estimate the correlation function using the \citet{Lan93} estimator
\begin{equation}
\xi=\frac{DD-2DR+RR}{RR} ,
\label{eq:LS}
\end{equation}
where DD, DR and RR are the suitably normalized numbers of weighted
data-data, data-random and random-random pairs in each separation
bin. We weight the galaxies (real and random) according to the angular
and radial selection functions. 
We use a simple weighting by the inverse of the selection function, as
the samples we use are all approximately volume-limited, and we have
verified that our results are insensitive to employing alternative 
weighting schemes. We also used the alternative $\xi$ estimators of
\citet{Dav83} and \citet{Ham93} and found no significant differences
in the results. 

Here, and throughout the paper, 
we estimate statistical errors on our measurements
using jackknife resampling with 104 angular subsamples.  Each subsample 
excludes roughly 37 square degrees (generally contiguous on the sky), 
which is about $90\hmpc$ comoving on a side at $z=0.3$.  
The 2.5 degree SDSS stripes are $36\hmpc$ comoving at $z=0.3$.
\citet{Zeh04b} performed extensive tests with mock catalogs to check the
reliability  of the jackknife error estimates over a similar range of
separations (see their Fig.~2). Their tests showed that the jackknife 
method is a robust way to estimate the error covariance matrix, especially 
for large volumes as probed here. 

\begin{figure}[tb] 
\plotone{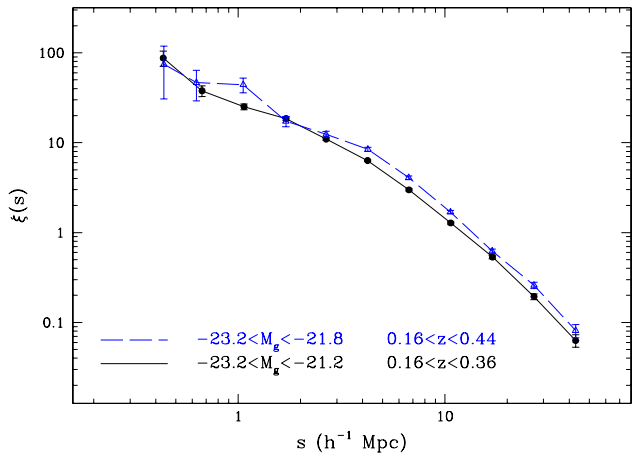}
\caption{\label{fig:xsis}
Redshift-space correlation function $\xi(s)$ for the LRG samples.
Bins in $s$ are in logarithmic separation of $0.2$. 
}
\end{figure}

Figure~\ref{fig:xsis}  shows the redshift-space correlation function,
$\xi(s)$, 
for the $-23.2<M_g<-21.2$ and $-23.2<M_g<-21.8$ LRG samples introduced in 
\S~\ref{subsec:samples}, with errorbars obtained from the jackknife 
resampling. 
The small difference in amplitude arises from the difference in the average
luminosity of the galaxies in the samples (see Table~\ref{tab:samples}), 
reflecting the known trend of stronger clustering with luminosity
(\eg, \citealt{hamilton88,park94,loveday95,benoist96,guzzo97,Nor01,Zeh02,Hog03,Zeh04b}).
The redshift-space correlation functions values are given in 
Table~\ref{tab:results}.

\begin{table*}[tb]\footnotesize
\caption{\label{tab:results}}
\begin{center}
{\sc Correlation Functions Measurements\\}
\begin{tabular}{cccccccccc}
\doubleline
\multicolumn{1}{c}{ } & \multicolumn{3}{c}{$-23.2<M_g<-21.2$} & \multicolumn{3}{c}{$-23.2<M_g<-21.8$} & \multicolumn{3}{c}{$-22.6<M_g<-21.6$} \\
separation & $\xi(s)$ & $\wrp$ & $\xi(r)$ & $\xi(s)$ & $\wrp$ & $\xi(r)$ & $\xi(s)$ & $\wrp$ & $\xi(r)$ \\
\singleline
0.418 & 87.4 (16.8) & 772.9 (52.7) & 675.7 (80.6) & 74.6 (44.0) & 1281 (243) & 1276 (399) & 39.6 (17.7) &  743.9 (86.8) & 484 (133) \\
0.663 & 37.7 (5.1) & 414.4 (26.0) & 210.3 (23.3) & 46.6 (17.5) & 575.2 (86.9) & 332.8 (84.8) & 30.3 (8.9) & 548.0 (66.8) & 316.9 (59.7) \\
1.051 & 25.1 (1.9) & 238.8 (13.4) & 66.9 (7.9) & 44.2 (8.3) & 286.8 (33.8) & 83.1 (21.8) & 30.9 (4.3) & 268.0 (29.0) & 75.5 (16.7) \\ 
1.665 & 18.5 (0.9) & 154.9 (7.5) & 24.9 (3.0) & 17.3 (2.3) & 178.5 (20.6) & 26.0 (7.8) & 19.4 (1.8) & 174.6 (14.1) & 30.8 (5.4) \\
2.639 & 11.0 (0.3) & 109.0 (5.1) & 11.7 (1.0) & 12.4 (1.0) & 132.4 (12.6) & 13.0 (3.2) & 12.1 (0.8) & 111.6 (9.0) & 10.1 (2.1) \\
4.182 & 6.32 (0.15) & 74.0 (3.8) & 5.04 (0.47) & 8.45 (0.45) & 95.5 (7.8) & 6.01 (1.11) & 7.35 (0.33) & 84.9 (7.1) & 5.58 (0.88) \\
6.628 & 2.99 (0.08) & 50.5 (2.6) & 2.32 (0.19) & 4.08 (0.19) & 70.8 (4.8) & 3.80 (0.50) & 3.52 (0.14) & 59.2 (4.1) & 2.62 (0.34) \\
10.505 & 1.28 (0.04) & 33.0 (2.2) & 1.04 (0.09) & 1.70 (0.08) & 38.8 (3.6) & 1.16 (0.19) & 1.52 (0.06) & 40.1 (3.1) & 1.25 (0.16) \\
16.650 & 0.54 (0.03) & 20.0 (1.8) & 0.43 (0.05) & 0.62 (0.03) & 25.2 (2.2) & 0.56 (0.09) & 0.59 (0.04) & 25.1 (2.5) & 0.57 (0.08) \\
26.388 & 0.19 (0.02) & 11.0 (1.4) & 0.14 (0.02) & 0.26 (0.02) & 13.2 (2.0) & 0.17 (0.06) &  0.23 (0.02) & 13.2 (1.8) & 0.20 (0.05) \\
\singleline
\end{tabular}
\end{center}
NOTES.---%
Measurements of the redshift-space correlation function, $\xi(s)$, projected 
correlation function, $\wrp$, and real-space correlation function, $\xi(r)$,
for the three LRG samples discussed in the paper. 
Correlation functions are calculated for each sample over the range for
which it is approximately volume-limited, denoted in Table~\ref{tab:samples}.
Comoving separations and $\wrp$ values are in $\hmpc$ units. Redshift-space 
$\xi(s)$ and real-space $\xi(r)$ are dimensionless. The diagonal terms
of the measurements error covariance matrices are given in parentheses.
Our radial bins are logarithmically spaced with widths of 0.2 dex
beginning at $10^{-0.49}$.  The separations listed in column 1 are the
linear centers of the bins.  Strictly speaking, the listed values of $\xi(s)$
and $w_p(r_p)$ are the averages of these correlation functions over the
annuli.  However, for reasonable power-law interpolations, the values
of $w_p(r_p)$ and $\xi(r)$ are very nearly ($\ll1\%$) the values at the
linear bin centers.  The interpolated value of $\xi(s)$ at the bin centers
would be about $1.5\%$ higher than the values in the table.
\end{table*}

\begin{figure}[tb] 
\plotone{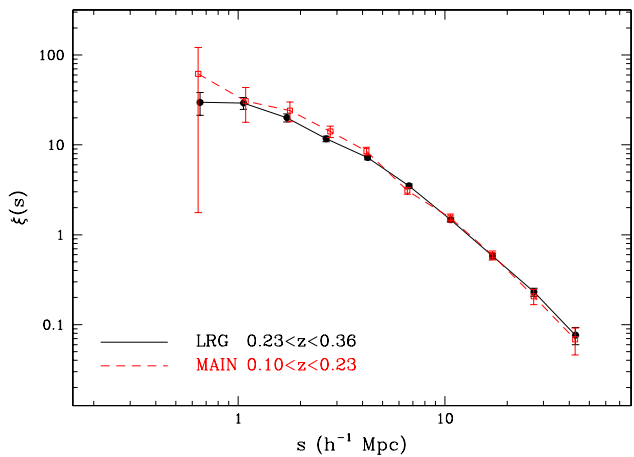}
\caption{\label{fig:xsislg}
Redshift-space correlation function $\xi(s)$ for the $-22.6<M_g<-21.6$ 
passively-evolved LRG sample and the comparable one from the MAIN galaxy 
sample.
}
\end{figure}

Figure~\ref{fig:xsislg} shows $\xi(s)$ for the $-22.6<M_g<-21.6$ LRG sample
plotted together with $\xi(s)$ obtained for the comparable red 
$-23<M_r<-22$ MAIN galaxy subsample of \citet{Zeh04b}, 
where we restrict the LRG sample to $0.23<z<0.36$, such that we are probing
independent volumes. The LRG
sample contains $\sim\!12400$ galaxies, while the MAIN galaxy sample 
includes only $\sim\!2700$. As is obvious from the plot, the agreement 
between the samples is excellent, and the LRG results thus extend in essence
the MAIN galaxy clustering results to higher redshifts. The deviations
at small separations are mainly due to shot noise effects arising from the
small number of galaxies in the MAIN sample and are consistent within
the errorbars.  The numerical values of this LRG correlation function
are also provided in Table~\ref{tab:results}.

\pagebreak

\section{Real-Space Clustering}
\label{sec:rclustering}

\subsection{Projected Correlation Function}
\label{subsec:projected}

To separate effects of redshift distortions from spatial correlations,
it is customary to estimate the correlation function on a two dimensional
grid of pair separations parallel ($\pi$) and perpendicular ($r_p$) to
the line of sight, termed $\xi(r_p,\pi)$. 
One can then learn about the real-space correlation function by
computing the projected correlation function
\begin{equation}
\wrp = 2 \int_0^{\infty} d\pi \, \xi(r_p,\pi). 
\label{eq:wp}
\end{equation}
In practice, we integrate up to $\pi_{max}=80\hmpc$, which is large enough
to include most correlated pairs and gives a stable result.
The omission of pairs at $\pi>80\hmpc$ likely causes an overestimate
of $w_p(r_p)$ by 1--$2\hmpc$, as the correlations on such scales are
driven negative by redshift distortions.  However, we have varied
$\pi_{max}$ from 50--120$\hmpc$ without significant change in $w_p$.
We also checked the robustness to binning in $r_p$ and in the integrated-over 
$\pi$ direction, finding the results to be insensitive to either.

The projected correlation function can in turn be related to the
real-space correlation function, $\xi(r)$,
\begin{equation}
\wrp= 2 \int_{r_p}^{\infty} r dr\; \xi(r)  (r^2-{r_p}^2)^{-1/2}
\label{eq:wp2}
\end{equation}
\citep{Dav83}.  
In particular, fitting a power-law to the $\wrp$ measurement allows
us to infer the best-fit power law for $\xi(r)$.

\begin{figure}[tb] 
\plotone{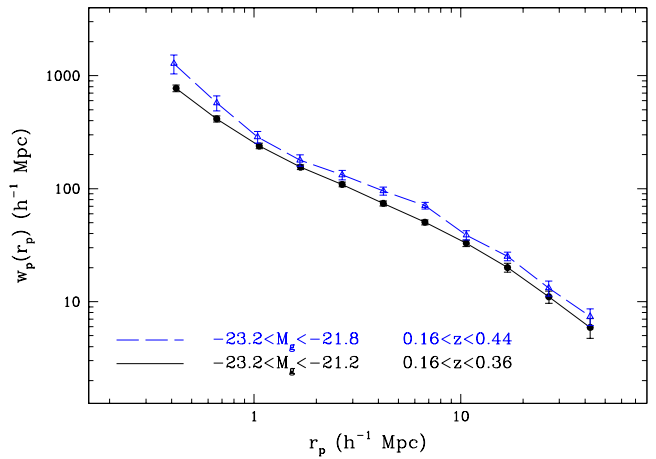}
\caption{\label{fig:wpl}
Projected correlation function $\wrp$ for the LRG samples.
}
\end{figure}

\begin{figure}[tb] 
\plotone{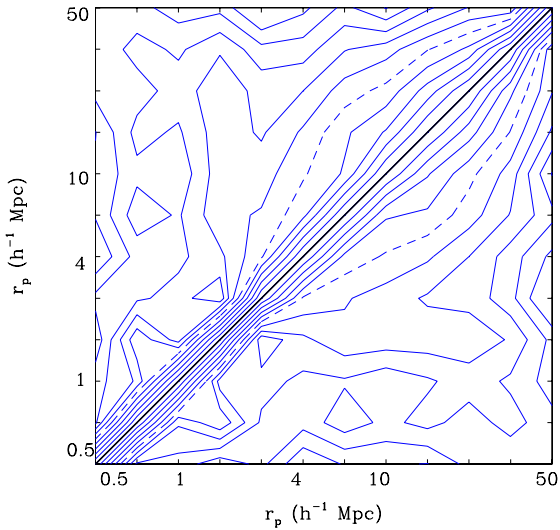}
\caption{\label{fig:covarn}
Normalized error covariance matrix for the $\wrp$ measurement of the
$-23.2<M_g<-21.2$ sample. The normalized covariance matrix is defined as
$C_{ij}/(C_{ii} \cdot C_{jj})^{1/2}$, where $C_{ij}$ are the elements of the
error covariance matrix. Contour spacing is 0.1 going from 1 on the 
diagonal (thick line) down to 0. The dashed lined denotes the 0.5 contour.
Tickmarks denote the elements in the covariance
matrix, and the labels denote the corresponding $r_p$ values.
}
\end{figure}

Figure~\ref{fig:wpl} shows the resulting projected correlation function,
$\wrp$, for the two inclusive LRG samples analyzed in this paper. Again,
the differences in amplitude reflect the luminosity bias between the samples. 
Figure~\ref{fig:covarn} shows the normalized (such that the diagonal
is 1) jackknife error covariance matrix of the $\wrp$ measurements
for the $-23.2<M_g<-21.2$ sample. As is apparent, there is
significant correlation between the measurements on different scales,  
but the auto-correlation along the diagonal is relatively
strong with the cross-correlation falling off rapidly.

\begin{figure}[tb] 
\plotone{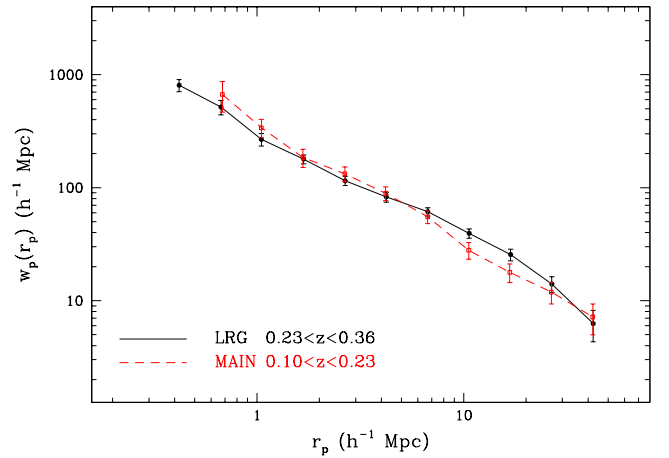}
\caption{\label{fig:wplg}
Projected  correlation function $\wrp$ for the $-22.6<M_g<-21.6$ 
LRG sample and the comparable one from the MAIN galaxy sample.
}
\end{figure}

Similar to the comparison to the MAIN galaxy sample results shown
in \S~\ref{sec:sclustering}, Figure~\ref{fig:wplg} compares the projected
correlation function of the analogous LRG and MAIN samples.
The small differences seen in the plot do not appear to be significant.
A $\chi^2$ statistic of the difference, performed with the sum of the error 
covariance matrices of the two measurements, is $11.6$
for the $10$ degrees of freedom, consistent with cosmic variance.

\begin{figure}[tb] 
\plotone{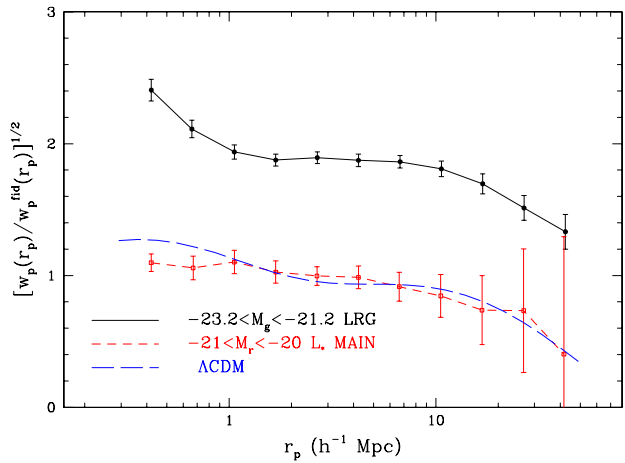}
\caption{\label{fig:bias}
$[\wrp/{w_p}^{fid}(r_p)]^{1/2}$ as a function of separation $r_p$ for the 
$-23.2<M_g<-21.2$ LRG sample (solid symbols and line) and for the 
$L_*$ MAIN galaxy sample (open symbols and short-dashed line;
\citealt{Zeh04b}), allowing to infer their relative bias.
${w_p}^{fid}(r_p)$ is the projected correlation function corresponding to 
a fiducial power-law
$\xi(r)=(r/5\hmpc)^{-1.8}$. The long-dashed curve shows this relative
quantity for a $\Lambda$CDM cosmology computed from the nonlinear power
spectrum of \citet{Smi03}.
}
\end{figure}

Now that we have demonstrated that the LRG sample extends the MAIN
galaxy sample, it is interesting to compare the amplitude of the LRG
clustering to that of typical $L_*$ galaxies. 
Figure~\ref{fig:bias} shows such a comparison for our largest 
LRG sample ($-23.2<M_g<-21.2$) and the $L_*$ MAIN galaxy sample
(a volume-limited sample with $-21<M_r<-20$ containing $5700$ galaxies; 
\citealt{Zeh04b}). Note that this is in contrast to the previous comparisons 
(Fig.~\ref{fig:xsislg} and \ref{fig:wplg}), where the LRG and red MAIN 
galaxy samples were chosen to match in luminosity and color.  
The quantities plotted are $[\wrp/{w_p}^{fid}(r_p)]^{1/2}$, where 
the fiducial ${w_p}^{fid}(r_p)$ corresponds to a power-law correlation 
function $\xi(r)=(r/5\hmpc)^{-1.8}$, and allow to infer the relative bias. 
For illustration purposes, we also plot 
this for a flat $\Lambda$CDM cosmology (with $\Omega_m=0.3$, 
$h=0.7$, $n=1$ and $\sigma_8=0.9$) projected correlation function
computed from the nonlinear power spectrum of \citet{Smi03} (Z.\ Zheng, 
private communication). The matter
correlation function is comparable in amplitude to the $L_*$ MAIN correlation
function, but distinct in detail. For the LRG galaxies, this scaled quantity 
appears roughly scale-invariant for $1-10\hmpc$,
with a notable upturn on smaller scales and a downturn on large scales.
When fitting a constant bias factor between the two samples, taking into
account the error covariance matrices, one obtains 
$b_{LRG}/b_* = 1.84 \pm 0.11$ when fitting over $1\hmpc < r_p \la 10 \hmpc$. 
The scale dependence of the LRG inferred bias is in accord with the
steeper correlation functions associated with red galaxies (\eg, 
\citealt{Wil98,Bro00,Zeh02,Zeh04b}).  The downturn of the projected 
correlation function from a power-law on large separations is similar to that 
predicted by CDM models and to what is measured in the SDSS MAIN galaxy 
sample and in the 2dF survey \citep{Haw03}.

\begin{figure}[tb] 
\plotone{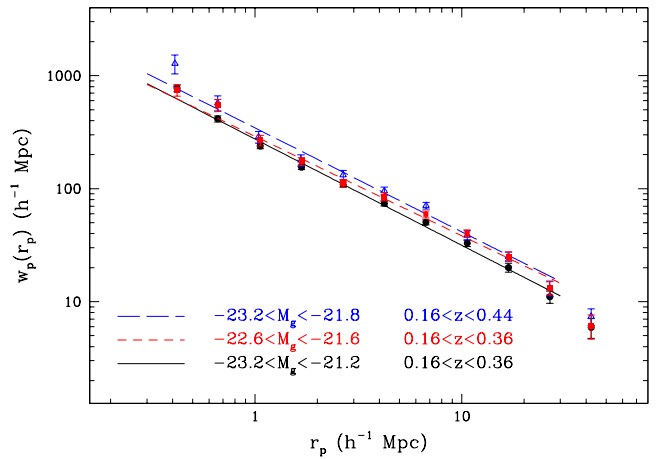}
\caption{\label{fig:wpl_pl}
Projected  correlation function $\wrp$ for the three LRG samples
discussed in the paper, plotted together with power-law
fits fitted over the range $0.3<r_p<30\hmpc$. 
}
\end{figure}

Real-space correlation functions have been historically well described
by power laws 
\citep{Tot69,Pee74,Got79,Dav83,Fis94,jing98,jing02,Nor01,Zeh02},
although recent precision measurements 
provide evidence for deviations from a power-law and a means of
explaining them (\eg, \citealt{Ber03,Mag03,Mal03,Zeh04a,Zhe04}).
Figure~\ref{fig:wpl_pl} shows power-law fits to our
projected correlation functions. The inferred $\xi(r)$ power-law fits 
are given in Table~\ref{tab:pl_fits}, while the $\wrp$ measurements themselves
are provided in Table~\ref{tab:results}. Inspection of the values of the 
correlation length, $r_0$, show clearly the trend with luminosity. The 
power-law slopes, $\gamma$, span the range 1.89 -- 1.94. 
The $\chi^2/d.o.f$ values for the power-law fits are in the range $2.3-3.9$,
indicating that a power-law is not a good fit. (The confidence level of a 
power-law fit is about 1.5\% in the best case and less than 0.1\% in the worst 
case.)

\begin{figure}[tb] 
\plotone{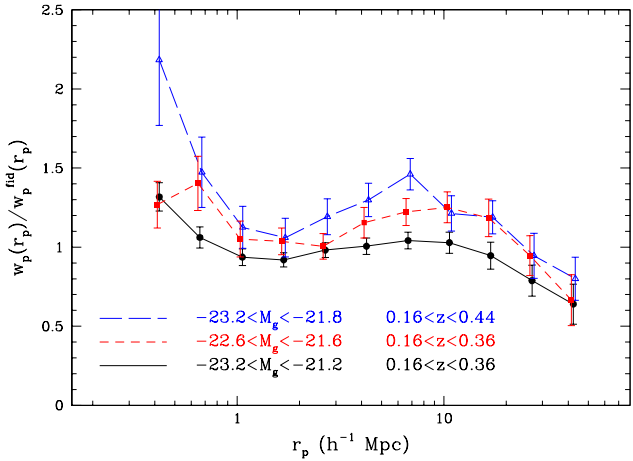}
\caption{\label{fig:wpl_relpl}
Projected  correlation function $\wrp$ for the LRG samples
shown in Fig.~\ref{fig:wpl_pl}, now all divided out by a fiducial
power-law $\wrp$ corresponding to $\xi(r)=(r/10\hmpc)^{-1.9}$. The deviations
from a power-law are clearly visible. 
}
\end{figure}

The deviations from a power-law are clearly visible in 
Figure~\ref{fig:wpl_relpl} where we divide the clustering measurements
by a representative power-law $\wrp$ corresponding to a $\xi(r)$ with 
$r_0=10\hmpc$ and $\gamma=1.9$. Similar deviations are also seen in the 
complementary analysis of the LRG samples by \citet{Eis04}.
These deviations appear to be of a similar nature to the deviations 
detected in the MAIN galaxy samples \citep{Zeh04a,Zeh04b}, which are naturally
explained by contemporary models of galaxy clustering as the transition
from a small-scale regime dominated by galaxy pairs in the same dark matter
halo to a large-scale regime dominated by pairs of galaxies in separate halos.
We delay to future work 
detailed halo modeling of this sort and interpretation of our measurements.  
There is a hint from the brightest sample of an increase at
small scales ($<1\hmpc$) in the luminosity dependence of the bias,
in agreement with the findings of \citet{Eis04}.

\begin{table}[tb]\footnotesize
\caption{\label{tab:pl_fits}}
\begin{center}
{\sc $\xi(r)$ Power Law Fits\\}
\begin{tabular}{cccc}
\doubleline
$M_g$ & $r_0$ & $\gamma$ & $\chi^2/d.o.f.$ \\
\singleline
$-23.2<M_g<-21.2$   &   $9.80  \pm 0.20$  &  $1.94 \pm 0.02$  &  $3.9$ \\
$-23.2<M_g<-21.8$   &   $11.21 \pm 0.24$  &  $1.92 \pm 0.03$  &  $3.1$ \\
$-22.6<M_g<-21.6$   &   $10.59 \pm 0.29$  &  $1.88 \pm 0.03$  &  $2.3$ \\
\singleline
\end{tabular}
\end{center}
NOTES.---%
$r_0$ and $\gamma$ are obtained from a fit to $\wrp$ using the full
error covariance matrix. The values of $r_0$ are given in $\hmpc$ units.
\end{table}

\vspace{0.5cm}
\subsection{Luminosity and Redshift Dependences}
\label{subsec:variations}

\begin{figure}[tb] 
\plotone{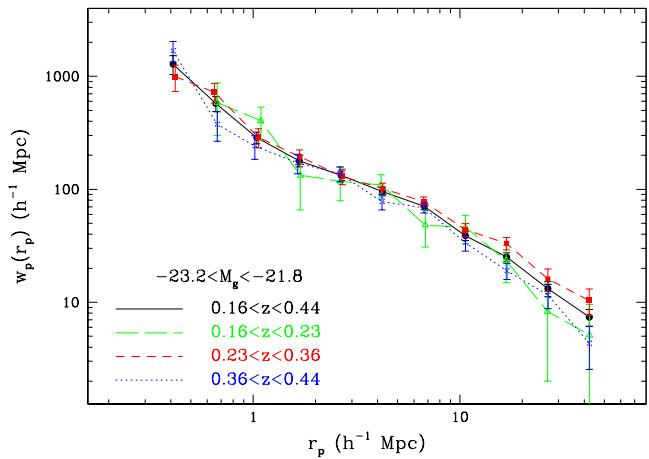}
\caption{\label{fig:z_dep}
Redshift dependence of the $\wrp$ clustering results for the 
$-23.2<M_g<-21.8$ sample.
}
\end{figure}

Figure~\ref{fig:z_dep} shows the redshift dependence of the
$-23.2<M_g<-21.8$ results. One can see small deviations of the results
corresponding to the different redshift ranges. For two independent redshift 
shells, we estimate the best-fitting multiplicative factor, $a$, between 
the two $\wrp$ measurements, taking into account the error covariance 
matrices. This factor would be significantly different than one if redshift 
evolution was present and consistent with one otherwise. 
The multiplicative factor between the $0.16<z<0.23$ and $0.23<z<0.36$ 
results is $a=0.84 \pm 0.14$  and between the $0.23<z<0.36$ and $0.36<z<0.44$ 
measurements it is $a=1.18 \pm 0.12$.
For our longest redshift baseline, we find $a=1.03 \pm 0.17$ between the
$0.16<z<0.23$ and $0.36<z<0.44$ measurements. Converting this to a limit
on $(1+z)^n$, we find $n=-0.2\pm1.1$.
We thus conclude that while some variations between the 
different redshift shells are present, these are likely to reflect large-scale 
structure variations, 
and that no consistent trend of redshift evolution is detected. We note that
for the $\wrp$ measurements for the comparable LRG and MAIN samples shown in 
Figure~\ref{fig:wplg}, $a= 1.01 \pm 0.10$, indicating clearly that no 
significant redshift evolution is present.

We also checked the robustness of our results when calculating
$\wrp$ separately for the two large disjoint areas of the northern Galactic
sky covered in the current SDSS samples.  The results from these two 
independent regions
are very similar and fully consistent, when calculated over the full 
redshift ranges of our samples. When looking at narrower redshift
shells the differences tend to be a bit larger, reflecting the slight 
variations with redshift seen in Figure~\ref{fig:z_dep}.  For example, the
tendency of $\wrp$ to have a slightly lower amplitude for the $0.16<z<0.23$ 
shell is reproduced 
in the off-equatorial region, while $\wrp$ for the equatorial region is 
similar to that of the full volume.
This supports our conclusion that these small deviations are sample
variance effects reflecting the large-scale structure fluctuations.

\begin{figure}[tb] 
\plotone{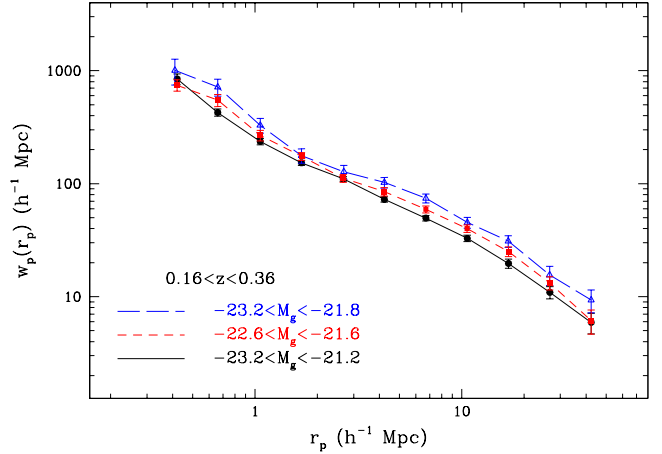}
\caption{\label{fig:L_dep}
Luminosity dependence of $\wrp$ obtained for the three LRG samples for
$0.16<z<0.36$.
}
\end{figure}

Figure~\ref{fig:L_dep} shows the projected correlation function $\wrp$
obtained for the three LRG samples over an identical volumes
($0.16<z<0.36$).
As mentioned previously, the small differences in clustering amplitude 
reflect the increase of clustering with luminosity. Again, we assess
the significance of the increased clustering amplitude by estimating the
best-fit multiplicative factor, $a$, between the measurements.  Since
these measurements are not fully independent (they are obtained from the 
same volume and thus are susceptible to similar cosmic variance effects),
we cannot treat their individual error covariance matrices as independent.
Instead, we estimate the value of $a$ from the mean and scatter of $a$ 
obtained from the individual jackknife realizations. The resulting
factor between the $-23.2<M_g<21.8$ and $-23.2<M_g<21.2$ measurements
is $a = 1.34 \pm 0.08$, more than a $4\sigma$ detection of luminosity
bias. For the $-22.6<M_g<-21.6$ versus the $-23.2<M_g<21.2$ measurements,
$a = 1.08 \pm 0.05$. It is clear that we detect a non-negligible luminosity
bias among the different LRG samples. 

\subsection{Real-Space Correlation Function}
\label{subsec:real-space}

\begin{figure}[tb] 
\plotone{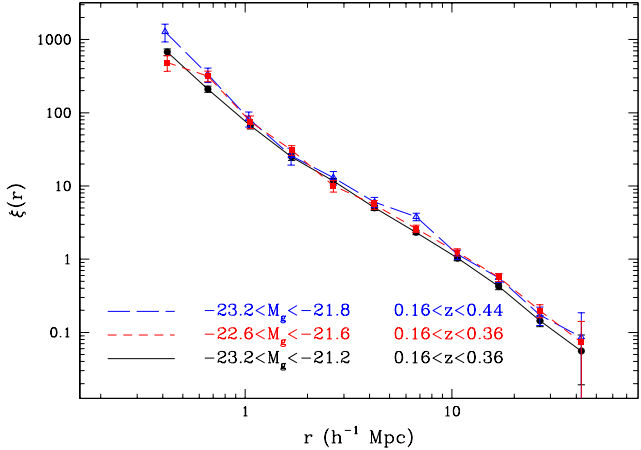}
\caption{\label{fig:xil}
Real-space correlation function $\xi(r)$ for the three LRG samples.
}
\end{figure}

It is possible to directly invert $\wrp$ to get $\xi(r)$ independent of 
the power-law assumption.  This is done by recasting Equation~\ref{eq:wp2} 
as
\begin{equation}
\label{eq:xi}
\xi(r) = - \frac{1}{\pi} \int_r^{\infty} dr_p \, {dw_p(r_p)\over{dr_p}} ({r_p}^2-r^2)^{-1/2}. 
\end{equation}
(\eg, \citealt{Dav83}).  We calculate the integral analytically
by linearly interpolating between the binned $\wrp$ values, following
\citet{Sau92}. We note that this estimate is only accurate to a few 
percent level, due to the inaccuracy of the linear interpolation.
Figure~\ref{fig:xil} presents the real-space correlation function, obtained
in this fashion, for the three LRG samples. The trends with luminosity and 
the hints of deviations from a power-law are noticeable here as well. 
The $\xi(r)$ values for these samples are given as well in 
Table~\ref{tab:results}.

It is common to summarize the amplitude of the correlation function
as the rms variation above Poisson
in the counts of galaxies in $R=8\hmpc$ comoving radius spheres.
The variance $\sigma_R^2$ can be calculated as
\begin{equation}
\sigma_R^2 = \int_{|\vec{r}_1|<R} d^3r_1 \
\int_{|\vec{r}_2|<R} d^3r_2 \xi(|\vec{r}_2-\vec{r}_1|)
\end{equation}
where the integrals are over the interior of two spheres of
radius $R$. This can be simplified to
\begin{equation}\label{eq:xi2sig}
\sigma_R^2 = \int_0^2 dy\;y^2\xi(yR)
\left( 3 -{9y\over 4} + {3y^3\over 16}\right),
\end{equation}
a useful formula that seems to have dropped out of the standard lore.
Following \citet{Eis03b}, we express this integral in terms of $w_p$ as
\begin{equation}\label{eq:wp2sig}
\sigma_R^2 = {1\over R^3} \int_0^\infty dr_p\;r_p w_p(r_p) g(r_p/R)
\end{equation}
where $g(x)$ is
$$
\left\{\begin{array}{ll}
{1\over 2\pi}\left[3\pi - 9x + x^3\right] & {\rm for\ } x\le 2, \\
{1\over 2\pi}\left[{-x^4+11 x^2 -28\over \sqrt{x^2-4}} + x^3 -9x + 6\sin^{-1}(2/x)\right] & {\rm for\ } x>2.
\end{array}\right.
$$
The kernel $g(x)$ is simpler than it looks: it starts positive, goes through
zero at $x\approx 1.28$, and then returns to zero as $-1/\pi x^3$ at large $x$.
It is differentiable at $x=2$.

We compute $\sigma_8$ by using Equation \ref{eq:wp2sig} and assuming
that $r_p w_p(r_p)$ is constant in each bin.  This yields
$\sigma_8 = 1.80 \pm 0.03$ for the $-23.2<M_g<-21.2$ sample and
$\sigma_8 = 2.06 \pm 0.05$ for the $-23.2<M_g<-21.8$ sample,
with the errors obtained from the jackknife subsamples.  We stress that this
is the real-space, non-linear $\sigma_8$, which should not be directly
compared with the linear-regime $\sigma_8$ values that are typically
quoted for CDM model normalization.

To facilitate comparison with the cross-correlations between
LRGs and $L^*$ galaxies presented by \citet{Eis04}, we also compute
the following integral of the real-space correlation function:
\begin{equation}\label{eq:Deltadef}
\Delta = {1\over V}\int_0^\infty 4\pi r^2dr\;\xi(r) W(r),
\end{equation}
where
\begin{equation}\label{eq:Wr}
W(r) = {r^2\over a_0^2} \exp\left(-{r^2\over 2a_0^2}\right),
\end{equation}
where $a_0$ is a constant scale factor.
The $\Delta$ statistic isolates about one octave of scale in the real-space
correlation function.
For power-law correlations of the observed slopes, 
$\Delta(a_0)\approx \xi(1.75a_0)$.  
The $\Delta$ values are computed as a simple non-singular
integral over $\wrp$.
For the $-23.2<M_g<-21.2$ sample, we compute
$\Delta = 66.5\pm3.5$, $15.6\pm0.7$, $4.40\pm0.19$, and $1.31\pm0.06$
for $a_0=0.5\hmpc$, $1\hmpc$, $2\hmpc$, and $4\hmpc$ {\it proper} distance
at $z=0.3$.  These scales are chosen to match those in 
\citet{Eis04}.  For the $-23.2<M_g<-21.8$ sample, the
$\Delta$ values are $91.8\pm9.6$, $18.1\pm1.8$, $5.54\pm0.43$, and 
$1.73\pm0.12$, respectively.

\section{Conclusions}
\label{sec:discussion}

We have presented a statistical analysis of the intermediate-scale 
($0.3$ to $40\hmpc$) correlations of $35,000$ luminous red galaxies 
from the SDSS, using three nearly volume-limited subsamples.  The size of 
the sample permits measurements of superb precision for these rare galaxies.  
We find clear deviations from power-law models in the projected correlation 
functions. Relative to a power-law, there are excesses of clustering on 
sub-Mpc scales and at $5-10$ Mpc scales.  These match qualitatively the 
deviations found in the SDSS MAIN sample galaxies and predicted by halo 
modeling \citep{Zeh04a,Zeh04b}.

The SDSS LRG sample reproduces the clustering results obtained with
the SDSS MAIN galaxy sample when matched appropriately in magnitude
and color and thus provides an extension of the galaxy clustering 
analyses to higher redshifts, and a better signal-to-noise measurement
of high luminosity galaxies.
We find no evidence for redshift evolution of the correlation functions 
(for a fixed passively evolving magnitude range) out 
to $z=0.4$.  However, the errors are such that dependences even as strong as
$(1+z)^{\pm2}$ cannot be excluded at more than $2\sigma$.

All of the LRG samples are highly clustered, with correlation lengths around 
$10\hmpc$ comoving, roughly twice that of $L_*$ galaxies 
(\eg, \citealt{Nor01,Zeh02,Zeh04b}). For the $-23.2<M_g<-21.2$ LRGs,  
the inferred bias relative to that of 
$L_*$ galaxies is $1.84 \pm 0.11$ on scales of $1 - 10 \hmpc$. 
The bias is roughly scale invariant on these scales and shows stronger 
clustering on smaller scales.

We find that more luminous LRGs are yet more clustered; however, none 
of our samples reach the correlation levels of rich clusters 
\citep{Bah83,Nic92,Pea92,Cro97,Aba98,Lee99,Col00,Gon02,Bah03}.
Note that all the latter estimate the cluster {\it redshift}-space correlation
function, and thus should be compared to our $s_0$ values inferred from 
Table~\ref{tab:results} and not the $r_0$ values quoted in 
Table~\ref{tab:pl_fits}.  The LRG clustering strengths and mean separations
$d$ are comparable to those of the poorest clusters mentioned in these works.
Our measurements are roughly consistent with the 
\citet{Bah03} trend of increasing correlation length with mean separation, 
$s_0 = 2.6 \sqrt{d}$ (see also their Fig.~2).
The LRG clustering strength we find is comparable as well to that of rich 
groups, which again have similar mean separations \citep{Pad04,Yan04}.

The SDSS LRG sample offers an enormous data set for the study of rare
but important massive early-type galaxies.
The interplay of number density, clustering amplitude, correlation function
shapes, redshift distortions, and higher-order correlations will provide a
rich data set for the modeling of the relationship of these galaxies
to their host halos and thereby to the evolution of the extreme end of the
galaxy mass function.

\acknowledgments
We thank Zheng Zheng for useful discussions and for providing the 
$\Lambda$CDM projected correlation function curve and Risa Wechsler
for useful comments. 
IZ and DJE are supported by grant AST-0098577 from the National Science
Foundation. DJE was further supported by an Alfred P.\ Sloan Research
Fellowship.

Funding for the creation and distribution of the SDSS Archive has been
provided by the Alfred P. Sloan Foundation, the Participating
Institutions, the National Aeronautics and Space Administration, the
National Science Foundation, the U.S. Department of Energy, the
Japanese Monbukagakusho, and the Max Planck Society. The SDSS Web site
is http://www.sdss.org/.

The SDSS is managed by the Astrophysical Research Consortium (ARC) for
the Participating Institutions. The Participating Institutions are The
University of Chicago, Fermilab, the Institute for Advanced Study, the
Japan Participation Group, The Johns Hopkins University, the Korean
Scientist Group, Los Alamos National Laboratory, the
Max-Planck-Institute for Astronomy (MPIA), the Max-Planck-Institute
for Astrophysics (MPA), New Mexico State University, University of
Pittsburgh, Princeton University, the United States Naval Observatory,
and the University of Washington.

\appendix
\section{Modeling of Selection Functions}
\label{sec:analysis}

\subsection{Redshift Distribution}

To perform a 3-dimensional correlation analysis, one must have a 
model for the expected number of galaxies (in the absence of clustering) 
as a function of redshift at every point on the sky.
Often this is done by computing a luminosity function and then
integrating to a given depth.  In the case of the LRG sample,
this is hard to implement reliably because the luminosity function is
so steep that minor variations in the $k$ and $e$ corrections 
are important and because the color selection does not select
the identical region of rest-frame color-luminosity space at
all redshifts.

Instead, we construct an approximate model of the redshift
distribution based on models of the selection and then low-pass
fit this model to the observed redshift distribution.
This removes power on very large radial separation scales
(below $k=0.04\ihmpc$), but provides an excellent model
for smaller scales.

In detail, 
we predict the $g-r$ and $r-i$ colors of early-type galaxies as a function
of redshift by convolving the observed average LRG spectrum \citep{Eis03a} with
the SDSS system response.  We then tweak that color-redshift relation onto
the observed photometry with low-pass filtering.  The spectral breaks
of early-type galaxies create subtle features in the color-redshift
relation that are currently difficult to resolve in the empirical photometry.
At each redshift, we create a Monte Carlo set of colors by adding Gaussian
noise to the mean color.  We then find how luminous a galaxy would have
to be to pass the LRG selection cut (this is a highly sensitive function
of color) and integrate a hypothesized passively-evolved luminosity function,
following \citet{Bla03c}, to find the number of galaxies.  Summing over
the Monte Carlo points and multiplying by the cosmological volume gives the 
predicted real density of galaxies per redshift bin for a passively-evolving 
model population.  This distribution is forced onto the observed redshift 
distribution using a boxcar smoothing length of 0.07 in redshift. Our 
resulting model is shown as the solid line in Figures~\ref{fig:nz210}, 
\ref{fig:nz216}, and \ref{fig:nz214}.
This more complex procedure is needed as, note, for example, the bump at 
$z=0.34$ that is caused by the G band absorption feature moving from the $g$
to $r$ filter.  A simple boxcar smoothing of the redshift histogram does
not preserve the height of this feature, but predicting the height from
the average LRG spectrum produces an excellent match.

\subsection{Angular Selection Function}

Selected galaxies can fail to receive spectra for three major reasons: 
1) they might fall within the fiber collision radius $55''$ of another 
target and not get assigned a fiber, 
2) the plate to which they were assigned might not have been
observed yet, and 3) they might have fallen outside of the boundary
of any plate.  
With the adaptive tiling of the survey, it is very rare for an isolated 
object inside a plate boundary to fail to be placed on the plate. 

We weight the LRGs to account for fiber collisions by using a
friends-of-friends grouping algorithm with a $55''$ linking length
(provided within the {\tt lss\_sample14} package) run on all of the
galaxy and quasar targets.  Within each collision group, we find the
number of objects that did get assigned to a plate and divide by the
total number.  The inverse of this is the weighting assigned to the
LRGs that did get assigned to plates.  Note that to the extent that the
selection of targets within the collision group was strictly random,
this approach is the correct procedure for two-point clustering on
separations larger than $55''$, even if the colliding objects are at
very different redshifts.  Imagine that there are two objects in close
proximity and we wish to study the correlations with a third distant
object (e.g., to count the pairs).  In a perfect survey, one would have
two pairs, each with a different redshift separation.  In one version
of the SDSS, one of the pairs would be counted, but with double weight;
in another version, the other pair would be counted with double weight.
The ensemble average over many such examples is unbiased with respect to
the result of the perfect survey.  Only the pair corresponding to the
two close objects is being mistreated (as it will never exist in this
example), but that pair only matters for very small separation structure.
In truth, the priorities of all targets are not quite equal, such that
LRGs will always lose to quasar candidates, but the priority is equal to 
the dominant MAIN targets. The key angular separation of $55''$ corresponds 
to $0.22\hmpc$ at $z=0.3$, and we thus restrict our measurements to 
separations larger than that.

The {\tt lss\_sample14} package provides the angular geometry of
the spectroscopic survey expressed in terms of spherical polygons.
The geometry is complicated: the spectroscopic plates are circular
and overlap, while the imaging is in long strips on the sky and 
there are some overlap regions of certain plates that may not have
been yet observed. The resulting spherical polygons track all these effects 
and characterize the geometry in terms of ``sectors'', each being a unique 
region of overlapping spectroscopic plates.
In each sector, we count the number of possible targets (LRG, MAIN, and
quasar), excluding those missed because of fiber collisions, and the number
of these that did get a spectrum.  
We weight the galaxies by the ratio of these numbers.
Typically, this ratio is very close to unity (mean of $1.007$).
It is large only in instances of regions covered by two plates where one of 
the plates has not yet been observed. To exclude these extreme cases, we 
cut our sample at a ratio of $1.67$, which eliminates $550$ LRG spectra.

We also mask from the sample any regions that are close to bright 
(heavily saturated) stars.
At present, we have not included any other masking.  Notably, we are
not explicitly excluding regions because of high reddening or poor
image quality; however, such regions are typically not targeted for
spectroscopy and would already have been excluded from the above geometry.
We are not masking out small-scale imaging defects such as saturation trails
or satellite tracks, nor avoiding the region around bright galaxies.
These involve less than $1\%$ of the survey area and have negligible effects
on the strong clustering signal shown by LRGs.
We treat the boundaries of the spectroscopic plates as simple boundaries
of the survey.  In other words, we neglect any correlation between these
real-world boundaries and the LRG distribution.  Since the LRGs are 
a subdominant population with mild angular clustering, it is very unlikely
that there is any important cross-talk between the LRG distribution 
and regions missed because of the adaptive tiling.

\subsection{Color Calibration Issues}

A persistent worry with clustering of SDSS LRGs is the high sensitivity
of the selection to errors in the photometric calibration \citep{Eis01}.  
Errors of 1\% in $r-i$ create 8\% fluctuations in the number density of the
$-23.2<M_g<-21.2$ sample, although the other samples are more robust 
because their
minimum luminosities are not close to the selection boundary and the
selection variations with absolute magnitude are three times less than
those with $r-i$ color.  Obviously, errors in the photometric zeropoints
will be correlated on the sky, leaving long stripes of false density
fluctuations on the sky.  

However, the SDSS has demonstrated 2\% rms calibrations \citep{Aba04}, 
which would produce only 0.025 effects in the correlation function.  
Moreover, we find that this
is an overestimate because the calibration errors are for the most part only
correlated along the SDSS stripes, not between them.  As the calibrations
are applied by secondary patches that are only about 0.6 degrees wide 
\citep{Sto02},
the calibration errors are suppressed on larger angular scales.  We see no
indication of any significant effect on the scales in this paper from calibration 
effects.

{}

\end{document}